\documentclass[twocolumn,aps,prl,superscriptaddress,showpacs,amsmath,amssymb,floats]{revtex4}

\usepackage{amsmath}
\usepackage{amssymb}
\usepackage{graphicx}
\usepackage{CJK}
\usepackage{keyval}
\usepackage{dcolumn}
\usepackage{bm}
\usepackage{color}
\usepackage{txfonts}
\usepackage[]{sidecap}

\begin{document}

\begin{CJK*}{UTF8}{gbsn}

\title{Anomalous correlation effects and unique phase diagram of electron doped FeSe revealed by angle resolved photoemission spectroscopy}

\author{C. H. P. Wen}\author{H. C. Xu}\author{C. Chen}\author{Z. C. Huang}\author{Y. J. Pu}\author{Q. Song}\author{B. P. Xie}
\affiliation{State Key Laboratory of Surface Physics, Department of Physics, and Advanced Materials Laboratory,
Fudan University, Shanghai 200433, People's Republic of China}

\author{Mahmoud Abdel-Hafiez}
\affiliation{Center for High Pressure Science and Technology Advanced Research, Shanghai 201203, China}

\author{D. A. Chareev}
\affiliation{Institute of Experimental Mineralogy, Russian Academy of Sciences, 142432 Chernogolovka, Moscow District, Russia}
\author{A. N. Vasiliev}

\affiliation{Low Temperature Physics and Superconductivity Department, M.V. Lomonosov Moscow State University, 119991 Moscow, Russia}

\author{R. Peng}\email{pengrui@fudan.edu.cn}

\author{D. L. Feng}\email{dlfeng@fudan.edu.cn}
\affiliation{State Key Laboratory of Surface Physics, Department of Physics, and Advanced Materials Laboratory,
Fudan University, Shanghai 200433, People's Republic of China}

\date{\today}
\begin{abstract}

In FeSe-derived superconductors, the lack of a systematic and clean control on the carrier concentration prevents the comprehensive understanding on the phase diagram and the interplay between different phases.
Here by K dosing and angle resolved photoemission study on thick FeSe films and FeSe$_{0.93}$S$_{0.07}$ bulk crystals, the phase diagram of FeSe as a function of electron doping is established, which is extraordinarily different from other Fe-based superconductors. The correlation strength remarkably increases with increasing doping, while an insulting phase emerges in the heavily overdoped regime. Between the nematic phase and the insulating phase, a dome of enhanced superconductivity is observed, with the maximum superconducting transition temperature of 44$\pm$2~K. The enhanced superconductivity is independent of the thickness of FeSe, indicating that it is intrinsic to FeSe. Our findings provide an ideal system with variable doping for understanding the different phases and rich physics in the FeSe family.

\end{abstract}


\maketitle

\end{CJK*}

The physical properties of correlated materials are sensitive to various parameters like carrier doping. Fine tuning on the carrier doping allows investigating the rich phase diagrams and helps understanding the interplay and mechanism of different phases.
However, for FeSe, a prototypical system of Fe-based superconductors with the simplest structure, the systematic doping control
is still lacking.
Although heavily electron doping has been achieved in intercalated FeSe crystals like $A_x$Fe$_{2-y}$Se$_2$ (\textit{A}=K, Rb, Cs, Tl/K) \cite{XLChen,KFeSe40K1} and (Li$_{0.8}$Fe$_{0.2}$)OHFeSe \cite{XFLu}, the doping levels are discrete and fixed.
Moreover, the phase separation in $A_x$Fe$_{2-y}$Se$_2$  \cite{Minghu1,Minghu2,FChen,JZhao,JQLi}complicates the studies on the intrinsic superconductivity.
In (Li$_{0.8}$Fe$_{0.2}$)OHFeSe, the polar surface prevents the observation of intrinsic bulk electronic structure in surface sensitive angle resolved photoemission spectroscopy (ARPES) measurements \cite{Niu}.
Single-layer FeSe films on SrTiO$_3$ or BaTiO$_3$ \cite{QKXue,XJZhou1,tan,XJZhou2,RPeng,RPengN} are another type of heavily electron doped FeSe-based superconductors, whose superconducting transition temperature (T$_{\mathrm{c}}$) could be above 65~K \cite{RPengN,JJF,mutual}. However, besides the electron doping, the interfacial effects are suggested as a crucial factor for the enhanced superconductivity \cite{ZX,RPengN}.
The lack of a clean FeSe system with systematic doping control prevents a full investigation on the phase diagram, and hampers the comprehensive understanding of the relationship between superconductivity and other ordered phases.

\begin{figure}[bt]
\includegraphics[width=8cm]{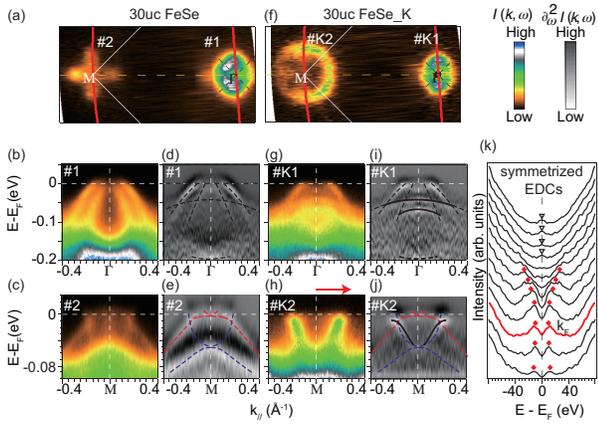}
\caption{(color online). (a) The photoemission intensity mapping at the Fermi energy ($E_F$) from a 30~uc FeSe film .
(b), (d) The photoemission intensity along cut \#1 in (a) and the corresponding second derivative, respectively.
(c), (e) The same as (b) and (d) but along cut \#2 in (a).
(f) The photoemission intensity mapping at $E_F$ from a 30~uc FeSe film with electron doping $x$=0.09 after K dosing.
(g), (i) The photoemission intensity along cut \#K1 in (a) and the corresponding second derivative, respectively.
(h), (j) The same as (g) and (i) but along cut \#K2 in (a).
(k) The symmetrized energy distribution curves (EDCs) along the momenta indicated by the arrows in panel (h). The data in (h), (j), and (k) were taken at 31~K, the others at 70~K.
}
\label{DopingEffect}
\end{figure}

\begin{figure*}[t]
\includegraphics[width=16cm]{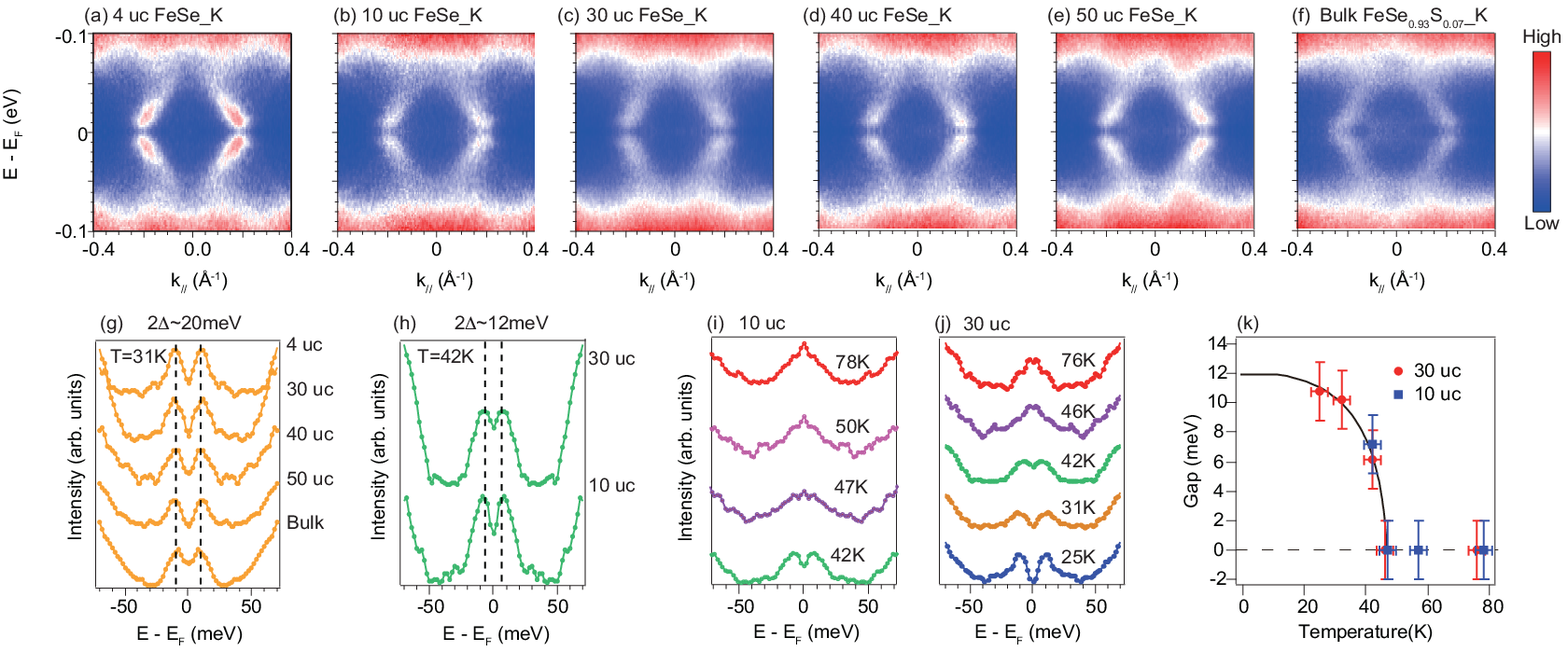}
\caption{(color online). (a)-(f) The symmetrized spectra of K-dosed FeSe films with thickness of 4~uc, 10~uc, 30~uc, 40~uc, 50~uc, and K-dosed bulk FeSe$_{0.93}$S$_{0.07}$, respectively. The data of 10~uc were taken at 42~K, the others at 31~K. (g) The symmetrized EDCs at k$_F$ for the FeSe films with different thicknesses and FeSe$_{0.93}$S$_{0.07}$ bulk crystal at 31~K after K dosing. (h) The symmetrized EDCs at k$_F$ for thicknesses of 10~uc and 30~uc at 42~K after K dosing. (i), (j) Temperature dependences of the symmetrized EDCs at k$_F$ for thicknesses of 10~uc and 30~uc, respectively. (k) The superconducting gap size as a function of temperature from the data in panels (i) and (j). The solid curve is the fitted result of BCS formula. The doping level is around 0.09. }
\label{ThickDep}
\end{figure*}

Recently, post-annealing in vacuum has been reported to effectively tune the doping in single-layer FeSe films on SrTiO$_3$ substrate, however, this approach fails in inducing superconductivity in the second layer FeSe \cite{XJZhou2,XJZhou3}.   Alternatively, by doping control with K dosing, a superconducting dome has been observed in FeSe films of 3~uc (unit cell) thickness \cite{Kdosed}. However, no superconductivity was found in 20~uc FeSe films at 13~K at any doping \cite{Kdosed}, therefore the enhanced superconductivity in 3uc FeSe/SrTiO$_3$ was attributed to some interface effect.

In this paper, we report the observation of an enhanced superconductivity in both thick FeSe films up to 50~uc and FeSe$_{0.93}$S$_{0.07}$ bulk crystals upon K dosing. The size of the superconducting gap and the T$_{\mathrm{c}}$ are identical on FeSe films with different thicknesses and on FeSe$_{0.93}$S$_{0.07}$ bulk crystals, indicating that the superconductivity is intrinsic to FeSe without any interfacial effects.  
More importantly, from the doping evolution of electronic structure and the superconducting behavior, a rich phase diagram of FeSe has been established, which show unique characteristics distinct from other Fe-based superconductors. We observe an anomalous enhancement of correlation strength with increasing doping. The dome of enhanced superconductivity with the highest T$_{\mathrm{c}}\sim$44$\pm$2~K is sandwiched between a nematic  phase and a correlation induced insulating phase. These results provide a global picture of the  interplay among nematic order, superconductivity, and electron correlations in the FeSe family.

The thick FeSe films were grown on TiO$_2$ terminated Nb:SrTiO$_3$ (001) substrates following the method described in our previous reports \cite{tan}. The electron doping is induced by depositing K atoms with a commercial SAES alkali dispenser. The doping levels no more than 0.158 were determined by ARPES based on Luttinger volume of Fermi surfaces, while the others were estimated according to the amount of K deposited. The single crystal of FeSe$_{0.93}$S$_{0.07}$ (T$_{\mathrm{c}}$=9.5~K) were grown using the flux method \cite{FeSeSgrowth,FeSeSgrowth2}.
ARPES data were taken under ultrahigh vacuum of 1.5$\times$10$^{-11}$~mbar, with a discharge lamp (21.2~eV He-I$\alpha$ light) and a Scienta~R4000 electron analyzer. The energy resolution is 7~meV and angular resolution is 0.3$^{\circ}$. The sample growth/cleaving, K deposition, and ARPES measurement  were all conducted in-situ.

\begin{figure*}[t]
\includegraphics[width=17cm]{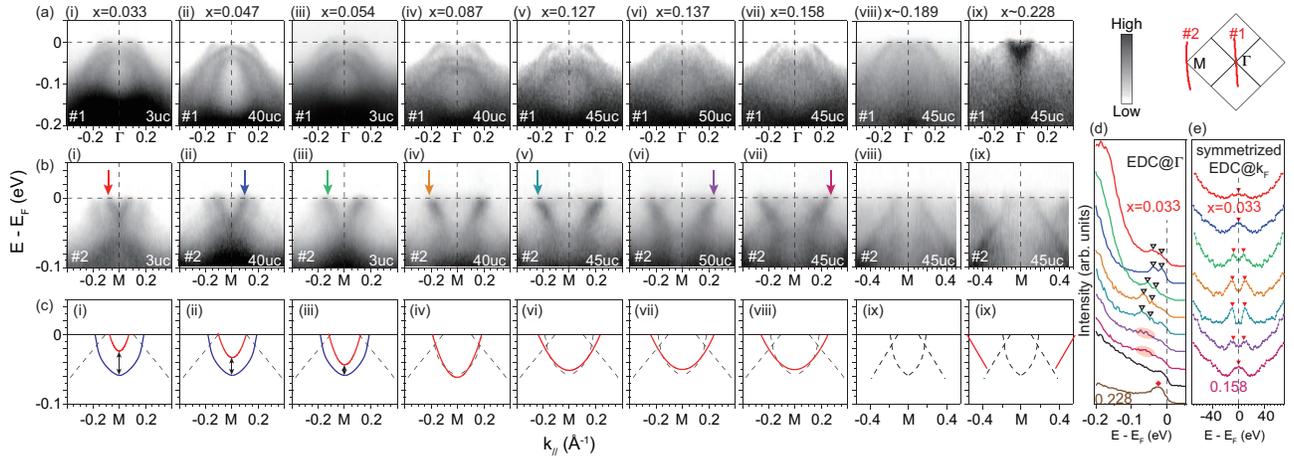}
\caption{(color online). (a) The evolution of photoemission spectra along cut \#1 as a function of increasing electron doping. (b) The doping dependent evolution of photoemission spectra along cut \#2. The upper-right inset shows the locations of cuts \#1 and \#2. (c) The doping dependent evolution of the  dispersions extracted from (b). The solid curves indicate the electron-like bands of the doped surface layer, while the dashed curves indicate  the dispersions from inner layers without doping.
(d) The EDCs at $\Gamma$ with different dopings.
(e) The symmetrized EDCs showing the evolution of superconducting gap as a function of doping. The momenta of spectra were indicated by the arrows in panels (b) with corresponding colors. The data in this figure were taken at 31~K, except those for x=0.054, which were taken at 25~K.}
\label{DopingDep}
\end{figure*}

Figure~\ref{DopingEffect} shows the band structures of a 30 unit cell (uc) thick FeSe before and after K dosing. Before K dosing, the band structures of the 30~uc FeSe film is consistent with those in the previous reports on thick FeSe films \cite{tan,ZYFeSe} and bulk FeSe crystals \cite{FeSebulk1,FeSebulk2,FeSebulk3,FeSebulk4}. As shown in Figs.~\ref{DopingEffect}(a), the Fermi surface consists of  hole pockets at $\Gamma$ and dumb bell shaped spectral weight at M. There are two hole-like bands crossing $E_F$ around $\Gamma$ (Fig.~\ref{DopingEffect}(b) and \ref{DopingEffect}(d)). Around M, the complex band structure is caused by  the splitting of bands with d$_{xz}$ and d$_{yz}$ orbital characters (Figs.~\ref{DopingEffect}(c) and \ref{DopingEffect}(e)) \cite{tan,ZYFeSe}, which reflects the orbital ordering or nematicity.
After K-dosing, a circular pocket appears around M [Figs.~\ref{DopingEffect}(f)]. The photoemission spectra show the superposition of two sets of band structures.  One set of bands follow the band structure of undoped FeSe and show weaker spectral weight, as indicated by dashed curves in Figs.~\ref{DopingEffect}(i) and \ref{DopingEffect}(j). Considering the finite detection depth of our ARPES measurement  \cite{tan},  these bands are attributed to the FeSe beneath the topmost layer. We found that the K atoms mainly dope the topmost unit cell, while the layers beneath remain undoped.
The second set of bands with the prominent photoemission spectral weight come from the topmost layer and are heavily electron doped. Around $\Gamma$, the two hole-like bands shift to higher binding energies and become flatter (Figs.~\ref{DopingEffect}(g) and \ref{DopingEffect}(i)). A simple electron-like band appears around M (Fig.~\ref{DopingEffect}(h) and \ref{DopingEffect}(j)), indicating that the nematic order is suppressed \cite{tan}. The carrier concentration is 9\% per Fe ($x$=0.09) according to the Fermi surface volume. Intriguingly,  the symmetrized energy distribution curves (EDCs) in Fig.~\ref{DopingEffect}(k) exhibit back bending  after passing the Fermi momentum (k$_F$) without crossing the Fermi energy. The sharp coherence peaks and back-bending behavior are hallmarks of Bogoliubov quasiparticle, which implie superconductivity in the K-dosed FeSe. The superconducting gap size is about 10~meV at 31~K, suggesting that the T$_{\mathrm{c}}$ in FeSe is significantly enhanced from the bulk T$_{\mathrm{c}}$ of 8~K. The weak features from the undoped inner layers remain gapless 
 around M [Fig.~\ref{DopingEffect}(k)], indicating that the superconductivity only exists in the doped topmost layer, without proximity into the  layers beneath. Our results are in contrast to the absence of superconductivity in 30~uc Fe$_{0.92}$Co$_{0.08}$Se thick films \cite{RPengN}, where the superconductivity is probably killed by the strong  scattering of the in-FeSe-plane Co ions  \cite{FeCoSe}. The enhanced superconductivity here suggest that the off-FeSe-plane K introduces  much weaker impurity scatterings \cite{ZRY}.

Figure~\ref{ThickDep} shows the thickness dependence of superconducting gap. At electron doping level around $x$=0.09,  back-bending dispersions and superconducting gaps are observed for all the K-dosed FeSe films  with  thicknesses varying from 4~uc to 50~uc [Figs.~\ref{ThickDep}(a)-\ref{ThickDep}(e)]. Moveover, for K-dosed FeSe$_{0.93}$S$_{0.07}$ bulk crystals with no FeSe/oxide interface, superconducting gap is also observed [Figs.~\ref{ThickDep}(f)]. At 31~K, the gap size  $\Delta$ is about 10~meV for all the films and bulk FeSe$_{0.93}$S$_{0.07}$ [Fig.~\ref{ThickDep}(g)].  For example, as shown in Fig.~\ref{ThickDep}(h), the gap size of 30~uc film is identical to that of 10~uc film at 42~K. 
With increasing temperature, the superconducting gap closes around 44$\pm$2~K for both films with thickness of 10~uc [Fig.~\ref{ThickDep}(i)] and 30~uc [Fig.~\ref{ThickDep}(j)]. Their temperature dependences are summarized in Fig.~\ref{ThickDep}(k), which can be well fit by the same Bardeen-Cooper-Schrieffer formula (BCS formula). Therefore, for thick films or bulk material,  the enhanced superconductivity here is intrinsic to electron doped FeSe, and does not dependent on the thickness or the FeSe/SrTiO$_3$ interface, which is distinct from the previous report on K-dosed FeSe \cite{Kdosed}.

The evolution of the electronic structure with electron doping is further studied by systematically  altering the K dosing. 
Figure ~\ref{DopingDep}(a) shows the spectra around $\Gamma$ as a function of doping. For all the spectra with different doping levels,  dispersions from the undoped FeSe layers underneath are always visible, which is independent of the doping  of the surface.  When the electron doping level x of the surface FeSe layer is increased from 0.033 to 0.127, two hole-like bands gradually shift to higher binding energy [Figs. ~\ref{DopingDep}(a) and~\ref{DopingDep}(d)].
 Simultaneously, these two bands become flat with x from 0.087 to 0.127~ [Figure ~\ref{DopingDep}(a)], then become incoherent for x$>$0.137, and disappears for x$\sim$0.189, indicating increasing correlation strength. From the EDCs at $\Gamma$ [Figure ~\ref{DopingDep}(d)], we can see that the two quasiparticle peaks turn into incoherent spectral weight (pink shadow) when x=0.137 and 0.158, and totally diminish when x reaches 0.189. After it is further doped to x$\sim$0.228, a small electron-like band emerges around the zone center [Figure ~\ref{DopingDep}(a)], and a well defined quasiparticle peak appears again [Figure ~\ref{DopingDep}(d)].

Around M, two electron-like bands are observed for the K-dosed FeSe with x=0.033 [Fig.~\ref{DopingDep}(b)], which are illustrated by the solid curves in Fig.~\ref{DopingDep}(c). Compared with the undoped band structure in Figs.~\ref{DopingEffect}(e), the upper band shifts downwards and the lower band remains at a fixed binding energy. Since the energy separation between them reflects the strength of the nematic order \cite{ZYFeSe}, the decreased energy separation with increasing doping indicates the weakening of nematicity. Eventually, these two bands become degenerate at x = 0.087, indicating the complete suppression of nematicity. As the doping further increases, the electron band gradually becomes flatter, indicating the enhanced correlation, consistent with the behaviors of  the bands around $\Gamma$.  Remarkably, when x$\sim$0.189, the spectral weight from the topmost layer depletes, while only the spectra from the inner layer remains around the zone corner. The absence of spectral weight for the K-dosed bands both around $\Gamma$ and M near $E_F$ indicates that FeSe turns into an insulating state with x$\sim$0.189. As the doping further increases to about 0.228, the topmost layer reenters a metallic state with a very large electron pocket. The data shown here were taken on four different samples with the thickness of 3uc, 40uc, 45uc, 50uc [noted in Figs.~\ref{DopingDep}(a) and ~\ref{DopingDep}(c)], respectively, and have been reproduced in another 6 samples. The band dispersions evolve in the same trend regardless of film thickness.

The symmetrized EDCs in Fig.~\ref{DopingDep}(e) give the doping dependence of the  superconducting gap. 
The gap opening is observed at the doping level 0.054, indicating a coexisting regime that the superconductivity is enhanced while the nematicity is not fully suppressed. The gap size increases to $\sim$10~meV at x=0.087, and does not change significantly from x=0.087 to x=0.127, and then decreases to 7~meV at x=0.137.  The gap closes for x=0.158, indicating that the T$_{\mathrm{c}}$ is below 31~K. The sample with high doping level around 0.228 is not superconducting at 31~K (see Supplementary Information). 

\begin{figure}[t]
\includegraphics[width=8.6cm]{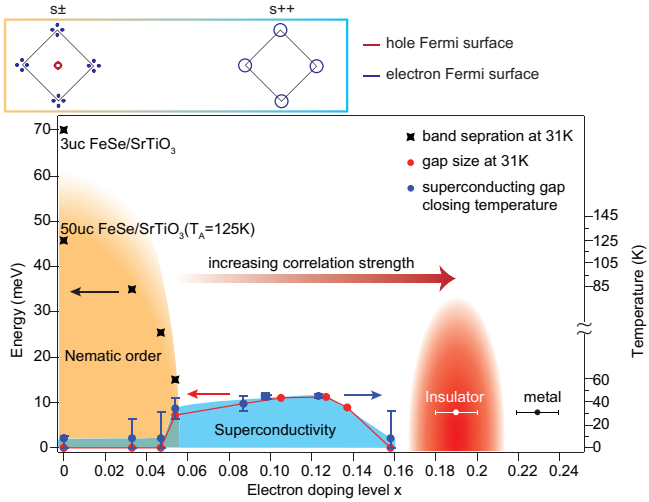}
\caption{(color online). Phase diagram of electron doped FeSe, and the summarize of the nematic band splitting, superconducting gap size, and the T$_{\mathrm{c}}$ as a function of doping. The nematic band splitting were determined by the energy difference between band bottoms in Fig.~3(c)(i-iv), while the undoped value is from ref.~\onlinecite{tan}. For doping values without generating superconducting gap at 31~K, T$_{\mathrm{c}}$'s were set as the T$_{\mathrm{c}}$ of bulk FeSe, 8~K. The other data points of T$_{\mathrm{c}}$ were determined by the superconducting gap-closing temperature. The gap size at 31~K were obtained by fitting the symmetrized spectra at $k_{\mathrm{F}}$. The upper inset illustrates the different Fermi surface topology and superconducting pairing symmetry of undoped FeSe and heavily electron doped FeSe.}
\label{PhaseDiagram}
\end{figure}

Figure~\ref{PhaseDiagram} summarizes the observed phases in K-dosed FeSe, and establishes a doping dependence phase diagram of K-dosed FeSe. By summarizing the superconducting gap size at 31~K, and the T$_{\mathrm{c}}$ determined by the gap-closing temperature (Supplementary Information), we have observed a superconducting dome with enhanced superconductivity near the nematic phase. The maximum T$_{\mathrm{c}}$ is 44$\pm$2~K, which is lower than the gap-closing temperature of 65~K in single-layer FeSe/SrTiO$_3$. The 21~K higher T$_c$ in single-layer FeSe/SrTiO$_3$ could be attributed to additional T$_c$ enhancement due to certain interface effect beyond carrier doping.

The phase diagram of electron doped FeSe has some essential ingredients of a canonical phase diagrams of iron based superconductors. For example, the superconductivity is enhanced when the nematic order is suppressed, and the superconductivity diminishes at high electron doping. However from the electronic structure perspective, it is actually rather distinctive from others and exhibits the following unique features.
\begin{enumerate}

\item  Both nematic order and superconductivity coexist in  undoped FeSe crystal  at low temperatures \cite{FeSebulk1,FeSebulk2,FeSebulk3,FeSebulk4}. The coexistence doping ranges from 0 to about 0.054, within which T$_{\mathrm{c}}$ even reaches  above 25 K.

\item  After the full suppression of nematicity,  the bandwidth narrows with  increased electron doping, indicating enhanced correlation. This is in contrast to many iron based superconductors such as LiFe$_{1-x}$Co$_x$As and 
NaFe$_{1-x}$Co$_x$As, where the bandwidth increases rapidly with electron doping \cite{ZRY}. 

\item  At the overdoped side with increasing electron doping, the system enters an insulating phase, which is quite extraordinary since most cuprates and iron-based superconductors become more Fermi-liquid-like and show decreasing correlation strength at the overdoped regime. This insulating phase is likely  a Mott insulator driven by the increased correlations. A second antiforromagnetic phase has been reported before in heavily electron doped 
LaFeAsO$_{1-x}$H$_x$ ($ x \sim$ 0.5) \cite{AF2}, and a superconductor-insulator transition has been reported in heavily electron doped (Li,Fe)OHFeSe through liquid gating \cite{gating}. The insulating phase discovered here is likely intimately related to those phases. In addition, an insulator to metal transition occurs with further electron doping in the far overdoped side.

\item  The Fermi surface of FeSe consists of hole pockets at $\Gamma$ and electron pockets around M \cite{tan,FeSebulk1,FeSebulk2,FeSebulk3,FeSebulk4}, where the superconducting paring symmetry is most likely to be s$_{\pm}$ type with sign reversal between the hole and electron pockets, as evidenced by previous experiments \cite{Hanaguri}. On the other hand, the Fermi surface of the electron doped FeSe consists of only electron pockets, where the pairing symmetry was found to be plain s-wave without any sign change for FeSe/STO \cite{QFanTZhang} and (Li$_{0.8}$Fe$_{0.2}$)OHFeSe \cite{Yan}. 

\end{enumerate}

To summarize, we obtained an enhanced superconductivity with T$_{\mathrm{c}}$ up to 44$\pm$2~K in thick FeSe films and FeSe$_{0.93}$S$_{0.07}$ bulk crystals by K-dosing. The superconductivity is independent on the film thickness and is intrinsic to K-dosed FeSe without the contribution from interface. 
Furthermore, we have observed a systematic evolution of electronic structure and established a unique phase diagram of FeSe with electron doping. A new insulating phase is observed at high doping levels, which is likely induced by increased correlation strength.  Our findings show that K-dosed FeSe can serve as a new and clean playground  with well-controlled electron doping and weak impurity scatterings for further studying  the relations between different phases, such as  the  evolution between  different pairing symmetries,  the superconductor-insulator transition, and the coexisting nematic order and superconductivity.

\textit{Acknowledgements:} We gratefully acknowledge Prof. J. P. Hu for helpful discussions. This work is supported in part by the National Science Foundation of China, and National Basic Research Program of China (973 Program) under the grant No. 2012CB921402.

\end{document}